\documentclass[letterpaper,11pt]{article}
\pdfoutput=1

\usepackage{jheppub}

\usepackage{multirow,subfig,xspace,xcolor}
\usepackage[countmax]{subfloat}

\newcommand{\zcut}{z_\text{cut}}

\newcommand{\rd}{\text{d}}

\DeclareRobustCommand{\Sec}[1]{Sec.~\ref{#1}}

\DeclareRobustCommand{\Fig}[1]{Fig.~\ref{#1}}

\DeclareRobustCommand{\Eq}[1]{Eq.~(\ref{#1})}

\DeclareRobustCommand{\Ref}[1]{Ref.~\cite{#1}}
\DeclareRobustCommand{\Refs}[1]{Refs.~\cite{#1}}

\allowdisplaybreaks

\title{Improving the Understanding of Jet Grooming in Perturbation Theory}

\author{Andrew J. Larkoski}
\affiliation{Physics Department, Reed College, Portland, OR 97202, USA}

\emailAdd{larkoski@reed.edu}

\abstract{
Jet grooming has emerged as a necessary and powerful tool in a precision jet physics program.  In this paper, we present three results on jet grooming in perturbation theory, focusing on heavy jet mass in $e^+e^-\to$ hadrons collisions, groomed with the modified mass drop tagger.  First, we calculate the analytic cross section at leading-order.  Second, using the leading-order result and numerical results through next-to-next-to-leading order, we show that cusps in the distribution on the interior of phase space at leading-order are softened at higher orders.  Finally, using analytic and numerical results, we show that terms that violate the  assumptions of the factorization theorem for groomed jet mass are numerically much smaller than expected from power counting.  These results provide important information regarding the convergence of perturbation theory for groomed jet observables and reliable estimates for residual uncertainties in a precision calculation.
}

\begin{document}
\maketitle

\section{Introduction}

A precision program for jet substructure calculations and measurements has developed through advances in jet grooming algorithms.  Because of its mitigation of non-global logarithms \cite{Dasgupta:2001sh} that would inhibit systematic improvability of theoretical predictions, the modified mass drop tagger (mMDT) groomer \cite{Dasgupta:2013ihk,Dasgupta:2013via}, and its generalization soft drop \cite{Larkoski:2014wba}, have emerged as the necessary tools for the precision task.  Following the original papers that introduced the groomers, a large literature of calculations and applications has resulted \cite{Frye:2016okc,Frye:2016aiz,Salam:2016yht,Marzani:2017mva,Marzani:2017kqd,Makris:2017arq,Hoang:2017kmk,Larkoski:2017iuy,Larkoski:2017cqq,Bendavid:2018nar,Kang:2018jwa,Chay:2018pvp,Baron:2018nfz,Makris:2018npl,Kang:2018vgn,Napoletano:2018ohv,Lee:2019lge,Marzani:2019evv,Gutierrez-Reyes:2019msa,Cal:2019gxa,Kang:2019prh,Chien:2019osu} and demonstrated that standard jet observables like the mass that have been groomed exhibit significantly improved sensitivity to the value of the strong coupling $\alpha_s$ and over a much wider dynamic range than its ungroomed counterpart.  This explosion of theoretical advances has been accompanied by measurements of groomed jet masses by both the ATLAS and CMS collaborations at the Large Hadron Collider (LHC) \cite{Aaboud:2017qwh,Sirunyan:2018xdh,Aad:2019vyi}.

For simplicity, much of these theoretical analyses have focused on jet production in $e^+e^-$ collisions, even further focused on center-of-mass energies of the $Z$ pole.  Recently, a re-analysis of archived data from the ALEPH experiment \cite{Badea:2019vey} at the Large Electron-Positron Collider (LEP) has demonstrated the proof-of-principle that studying jet grooming in $e^+e^-$ collisions can be more than just a purely academic exercise.  In this paper, we restrict to jets in $e^+e^-$ collisions for these reasons.  A precision prediction of any event or jet shape at a lepton collider requires three broad components: fixed-order calculations in the perturbation theory of QCD, resummation of large logarithms near the exclusive phase space boundaries to all orders in the coupling, and the dominant corrections from non-perturbative physics in the bulk of the phase space.  Advances have been made in all three of these directions for mMDT grooming in particular.  Next-to-next-to-leading order (NNLO) predictions for groomed jet mass has been computed \cite{Kardos:2018kth} in the CoLoRFulNNLO subtraction method \cite{DelDuca:2016csb,DelDuca:2016ily,Kardos:2016pic}.  Using the factorization theorem of \Refs{Frye:2016okc,Frye:2016aiz}, supplemented with two- and three-loop results \cite{Bell:2018vaa,Bell:2018oqa,Kardos:2020ppl,Bell:2020yzz}, next-to-next-to-next-to-leading logarithmic order (NNNLL) resummed predictions have been presented \cite{Kardos:2020gty}.  In \Ref{Hoang:2019ceu}, the first matrix element definition of non-perturbative corrections was provided for these groomers, with the leading contributions encapsulated into three universal coefficients.  Through appropriate combination of these results, predictions for mMDT jets in $e^+e^-$ collisions can be provided that rival the precision established of classic event shapes such as thrust \cite{Becher:2008cf,Abbate:2010xh} and $C$-parameter \cite{Hoang:2014wka,Hoang:2015hka}.

However, even in restricting analysis to mMDT groomed jets in $e^+e^-$ collisions, there are as of yet unresolved issues with the precision predictions that have been presented.  In addition to the scale enforced by the measurement of the jet mass, the groomer introduces another scale that defines which emissions are kept or removed from a jet.  The measurement scale and the grooming scale play off one another and result in interesting structure in the resulting distribution, depending on the relative size of these two scales.  Where the value of the jet mass is equal to the grooming scale, the leading-order distribution develops a cusp, and this may lead to significantly inaccurate higher fixed-order predictions in the vicinity \cite{Catani:1997xc}.  The factorization theorem of \Refs{Frye:2016okc,Frye:2016aiz} is only valid when the grooming scale is parametrically larger than the jet mass, but this isn't necessarily the regime that is most relevant for experiment.  The numerical size of corrections to the factorization theorem description hasn't been firmly established, which calls into question its relevance as the dominant description of the groomed jet near the exclusive phase space boundary.

In this paper, we address these issues directly and establish that their effect is actually substantially numerically smaller than would be na\"ively expected.  
In \Sec{sec:lo} we present the analytic prediction of the leading-order distribution of the groomed heavy hemisphere mass, which provides a foundation for the analyses in the following sections.  In \Sec{sec:cusp}, we study the cusp in the leading-order distribution of the groomed heavy hemisphere mass and show explicitly using numerical next-to- and next-to-next-to-leading order codes that the cusp is softened, contrary to what one might expect.  In \Sec{sec:factvio}, using numerical fixed-order codes, we isolate the contribution to the groomed heavy hemisphere mass distribution that is not described by the factorization theorem and show that its numerical size is about a factor of 4 times smaller than would be expected, for experimentally-relevant values of the grooming parameter.  We conclude and discuss future directions in \Sec{sec:concs}.

\section{Leading-Order Distribution}\label{sec:lo}

As mentioned in the introduction, we restrict our attention to jets produced in $e^+e^-$ collisions, which requires a slightly modified definition of the mMDT groomer than that presented in its original form \cite{Dasgupta:2013ihk}.  As $e^+e^-$ collisions occur in the center-of-mass frame, we groom each event hemisphere individually.  Once the events have been groomed, we then measure the masses of the event hemispheres.  Grooming decorrelates the hemispheres, and so a more natural scale to compare the mass to is the ungroomed hemisphere energy, rather than the center-of-mass energy.\footnote{The groomed mass must be compared to the ungroomed energy because the groomed jet energy is not infrared and collinear safe \cite{Baron:2018nfz}.}  We then only measure the ``heaviest'' of the two hemisphere masses.  Details of the precise algorithm can be found in, e.g., \Ref{Kardos:2020ppl}.

With this definition of our measurement procedure, it is straightforward to analytically calculate the leading-order distribution for the heavy hemisphere groomed jet mass $\rho$.  We first note that at leading order in the center-of-mass frame, one event hemisphere has two particles in it, while the other has only a single particle.  Thus, the heaviest hemisphere must be the one with two particles.  Using the three-body phase space variables $\{x_i\}$, where
\begin{equation}
x_i = \frac{2p_i\cdot Q}{Q^2}\,,
\end{equation}
where $Q$ is the total four-momentum of the event and $i=1,2,3$ ranges over the final state particles, the energy of the hemisphere with two particles is:
\begin{equation}
E_\text{heavy} = (2-\max\{x_i\})\frac{\sqrt{Q^2}}{2}\,.
\end{equation}
The least energetic particle of the event is also the least energetic particle of the two-particle hemisphere, with energy
\begin{equation}
E_\text{lo} = \min\{x_i\}\frac{\sqrt{Q^2}}{2}\,.
\end{equation}
The mMDT grooming requirement on the heavy hemisphere enforces that the groomed mass is only non-zero if
\begin{equation}
\frac{E_\text{lo}}{E_\text{heavy}} = \frac{\min\{x_i\}}{2-\max\{x_i\}} > \zcut\,.
\end{equation}

If the grooming requirement is satisfied, then the groomed jet mass is just the total hemisphere mass:
\begin{equation}
m^{(g)} = m^{(\text{heavy})} = \sqrt{1-\max\{x_i\}}\,\sqrt{Q^2}\,,
\end{equation}
in terms of the three-body phase space variables.  The observable of interest $\rho$ is then the ratio of this mass to the hemisphere energy:
\begin{equation}
\rho = \left(
\frac{m^{(\text{heavy})}}{E_\text{heavy}}
\right)^2 = \frac{4(1-\max\{x_i\})}{(2-\max\{x_i\})^2}\,.
\end{equation}
The leading-order distribution of $\rho$ can then be calculated from integrating over the matrix element for $e^+e^-\to q\bar q g$ production:
\begin{align}
\frac{1}{\sigma_0}\frac{d\sigma^{(0)}}{d\rho} &= \frac{\alpha_s C_F}{2\pi}\int_0^1dx_1 \int_0^1 dx_2\,\Theta(x_1+x_2-1)\frac{x_1^2+x_2^2}{(1-x_1)(1-x_2)}\\
&
\hspace{4cm}
\times\delta\left(
\rho - \frac{4(1-\max\{x_i\})}{(2-\max\{x_i\})^2}
\right)\Theta\left(
\frac{\min\{x_i\}}{2-\max\{x_i\}}-\zcut
\right)\,,\nonumber
\end{align}
where $\sigma_0$ is the leading-order electroweak cross section for $e^+e^-\to q\bar q$ and $C_F=4/3$ is the fundamental Casimir of SU(3) color.

\begin{figure}[t]
\begin{center}
\includegraphics[width=10cm]{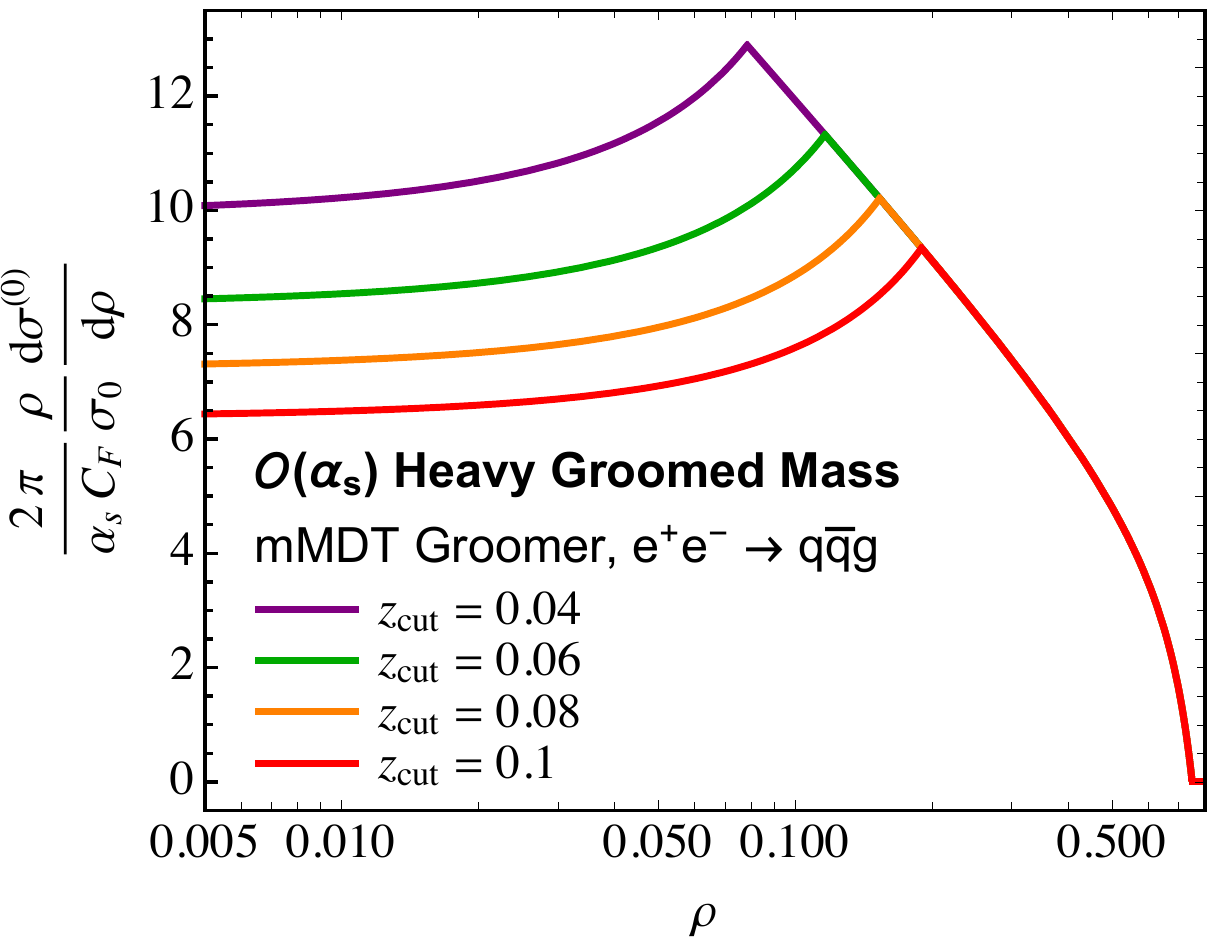}
\caption{Plots of the leading-order distribution of the groomed heavy hemisphere mass $\rho$ in $e^+e^-$ collisions, for values of the grooming parameter $\zcut = 0.04,\,0.06,\,0.08,\,0.1$.  The cusp in these distributions lives at $\rho = 2\zcut - \zcut^2$.
\label{fig:lo}}
\end{center}
\end{figure}

The phase space constraints are simple enough that the integral can be evaluated exactly.  We find
\begin{align}
\frac{2\pi}{\alpha_s C_F}\frac{1}{\sigma_0}\frac{d\sigma^{(0)}}{d\rho} &=\Theta\left(
\frac{3}{4}-\rho
\right)\Theta\left(
\rho-(2\zcut-\zcut^2)
\right)\left[
-\frac{12\left(
6-6\sqrt{1-\rho}+\rho(-8+5\sqrt{1-\rho}+2\rho)
\right)}{\rho^3(1-\rho)}\right.\nonumber\\
&
\hspace{1cm}
\left.
-\frac{2\left(6-6\sqrt{1-\rho}-\rho(5-4\sqrt{1-\rho})\right)}{\rho^2(1-\rho)}\,\log\frac{\rho}{2+2\sqrt{1-\rho}-3\rho}
\right]\\
&
\hspace{0.5cm}
+\Theta(2\zcut-\zcut^2-\rho)\left[
\frac{12(1-2\zcut)\left(2-2\sqrt{1-\rho}-\rho\right)^2}{\rho^3\left(2-2\sqrt{1-\rho}-\rho(2-\sqrt{1-\rho})\right)}\right.\nonumber\\
&
\hspace{-1cm}
\left.
-\frac{2\left(6-6\sqrt{1-\rho}-\rho(5-4\sqrt{1-\rho})\right)}{\rho^2(1-\rho)}\,\log\frac{2-4\zcut(1-\zcut-\sqrt{1-\rho})-2\sqrt{1-\rho}-\rho}{4\zcut(1-\zcut)-\rho}
\right]\,.\nonumber
\end{align}
This distribution is plotted in \Fig{fig:lo} for a few values of the grooming parameter $\zcut$.  The cusp in the distribution located at $\rho = 2\zcut - \zcut^2$ is clear: for values of $\rho$ above the cusp, grooming has no effect, while for $\rho$ below the cusp, grooming significantly modifies the distribution from its ungroomed counterpart.  

With an analytic result, it is interesting to isolate components of the distribution in different limits.  First, in the limit that $\rho \ll \zcut$, but $\zcut$ is arbitrary, the cross section reduces to
\begin{align}\label{eq:lolp}
\frac{\rho}{\sigma_0}\frac{d\sigma^{{(0)}\rho\ll\zcut}}{d\rho} =\frac{\alpha_s C_F}{2\pi} \left(
-3-4\log\zcut+6\zcut+4\log(1-\zcut)\right)
\,.
\end{align}
Thus, in this limit, this logarithmic cross section approaches a constant value, set by the value of $\zcut$.  Additionally, the first two terms in the parentheses on the right, $-3-4\log\zcut$, survive in the $\zcut \ll 1$ limit.  This sequential strongly-ordered limit $\rho \ll \zcut \ll 1$ is that described by the factorization theorem of \Refs{Frye:2016okc,Frye:2016aiz}.  The terms relevant for $\zcut\sim 1$, $6\zcut+4\log(1-\zcut)$, have not yet been calculated to arbitrary accuracy within a factorization theorem.  These terms arise from collinear splittings at leading power in $\rho\ll 1$, because soft, wide-angle emissions that pass the groomer enforce that $\zcut \ll 1$.  They were first calculated explicitly in \Ref{Marzani:2017mva}, which incorporated finite $\zcut$ effects into resummation of groomed mass for narrow jets at next-to-leading logarithm, following a proposal from the original paper on the mMDT groomer \cite{Dasgupta:2013ihk}.

We can also isolate the distribution around the cusp with weak grooming, where $\rho \sim \zcut \ll 1$.  In this region, the cross section becomes
\begin{align}
\frac{\rho}{\sigma_0}\frac{d\sigma^{{(0)}\rho\sim\zcut\ll1}}{d\rho} &= \frac{\alpha_s C_F}{2\pi}\left[\Theta(\rho-2\zcut)\left(
-3-4\log\frac{\rho}{2}+4\log\, 2
\right)\right.\\
&
\hspace{2cm}\left.+\,\Theta(2\zcut-\rho)\left(
-3-4\log\zcut-4\log\left(
1-\frac{\rho}{4\zcut}
\right)\right)
\right]\,.\nonumber
\end{align}
This expression is continuous through $\rho = 2\zcut$, but not smooth, which can be verified by differentiating above and below $\rho = 2\zcut$.  Just above $\rho=2\zcut$ we have
\begin{align}
\left.\rho\frac{d}{d\rho}\left(
-3-4\log\frac{\rho}{2}+4\log\, 2
\right)\right|_{\rho=2\zcut} = -4\,,
\end{align}
while just below $\rho = \zcut$ we find
\begin{align}
\left.\rho\frac{d}{d\rho}\left(
-3-4\log\zcut-4\log\left(
1-\frac{\rho}{4\zcut}
\right)\right)\right|_{\rho=2\zcut}=4\,.
\end{align}
Thus at leading power in $\zcut$ only the position of the cusp depends on $\zcut$, but not its shape, as also seen in \Fig{fig:lo}.  We will identify more features of this cusp in the following section.

\section{Cusps at Fixed Order}\label{sec:cusp}

With the analytic result for the leading-order cross section established in the previous section, we can calculate the discontinuity of the derivative of the leading-order cross section at the point where $\rho = 2\zcut-\zcut^2$, for arbitrary $\zcut$.  The difference in the derivative above and below that point is
\begin{align}
\frac{1}{\sigma_0}\frac{d}{d\log\rho}\left[
\frac{d\sigma^{(0)+}}{d\log\rho}-
\frac{d\sigma^{(0)-}}{d\log\rho}
\right]_{\rho = 2\zcut-\zcut^2} = - \frac{\alpha_s C_F}{2\pi}\frac{16(2-6\zcut+6\zcut^2-\zcut^3)}{(1-\zcut)^2(2-\zcut)(2-3\zcut)}\,,
\end{align}
where the $+$ and $-$ superscripts denote above and below the point $\rho  = 2\zcut - \zcut^2$, respectively.  As $\zcut\to 0$, this reduces to the difference calculated in the previous section.

A cusp located on the interior of phase space in a differential distribution can potentially produce unreliable predictions at higher fixed orders \cite{Catani:1997xc}.  These are typically caused by end points in low-order distributions that are not at the edge of the full phase space.  The cusp introduces a new ``boundary'' of phase space at that point at which the derivative of the cross section is discontinuous.  At higher orders, points immediately below the cusp can correspond to a degenerate phase space configuration in which virtual corrections are added to the leading-order prediction.  Points immediately above the cusp can be generated by soft or collinear real emissions off of the leading-order configuration.  Thus, immediately above and below the cusp, there can be a mis-cancelation of real and virtual divergences in the derivative of the cross section.  The differential cross section itself can still be continuous, but further higher-order corrections can transform the cusp to become more and more step-like.  This feature is observed, for example, at the endpoint of the leading-order distribution of thrust, where $\tau = 1/3$.  The next-to-leading order correction extends beyond $\tau = 1/3$, but begins to form a step-like shape around $\tau = 1/3$ \cite{Catani:1996jh}.  

The general analysis of \Ref{Catani:1997xc} would seem to suggest that the cusp observed in the groomed heavy hemisphere mass distribution would transform into a discontinuous step with the inclusion of higher fixed-order contributions.  Unlike the examples studied in that paper, though, the cusp in the groomed mass distribution lives on the interior of the phase space even at leading-order, so its higher-order corrections will have a different structure than, say, the $\tau=1/3$ end point in thrust.  If it were the case that this groomed cusp developed into a step, then the fixed order expansion would not smoothly converge around $\rho = 2\zcut-\zcut^2$, and this could be problematic for claiming theoretical precision throughout the distribution.  While no evidence for such a step has been observed in studies of mMDT grooming at next-to-leading order and beyond \cite{Frye:2016aiz,Baron:2018nfz,Kardos:2018kth}, this could simply be due to the fact that these studies used relatively large grid spacing in $\rho$ for the numerical fixed-order results.  The immediate region around the cusp hasn't been studied with sufficient resolution to identify step-like behavior or not at higher-orders.

\begin{figure}[t]
\begin{center}
\includegraphics[width=10cm]{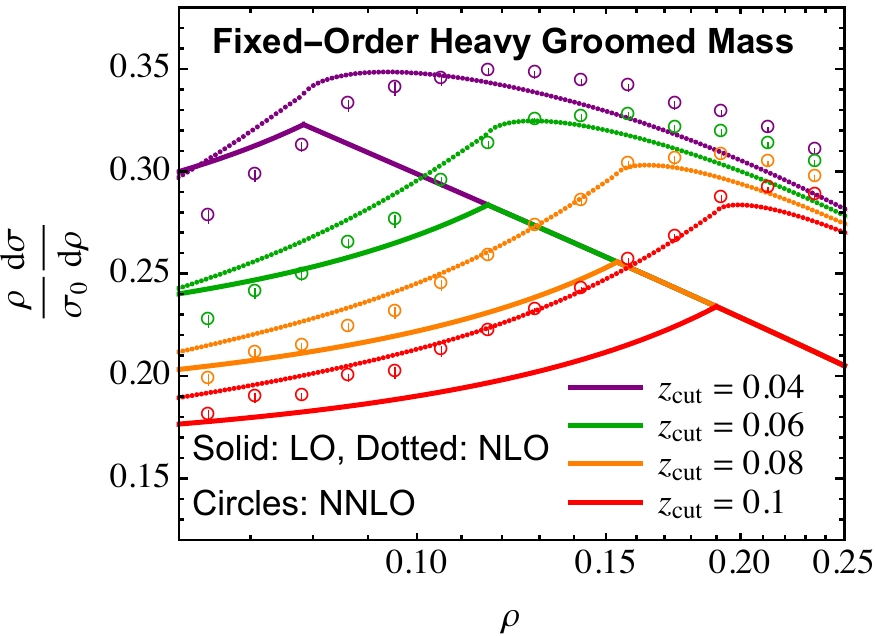}
\caption{Plots of the leading, next-to-leading and next-to-next-to-leading fixed order cross sections of the heavy hemisphere groomed mass, with $\zcut = 0.04\,, 0.06\,, 0.08\,, 0.1$.  These plots focus around the location of the cusp at leading order where $\rho_\text{cusp} = 2\zcut -\zcut^2$ and we have set $\alpha_s = 0.118$.
\label{fig:cusp}}
\end{center}
\end{figure}

To study the higher-order behavior of the cusp in the groomed heavy hemisphere distribution, we use results from fixed-order codes.  At next-to-leading order, we generated $10^{13}$ events at next-to-leading order in EVENT2 \cite{Catani:1996vz}, with grooming parameter $\zcut = 0.04\,, 0.06\,, 0.08\,, 0.1$.  From these events, we generated histograms with 400 uniform bins in $\log\rho$ in the range $\log\rho \in[-4,0]$.  This range is sufficient to cover the location of the cusp for each value of $\zcut$ considered and the bins are small enough to clearly resolve the cusps.  At next-to-next-to-leading order, we use the results generated with the CoLoRFulNNLO method, originally for the study of \Ref{Kardos:2020ppl}.  Details about event generation can be found in that reference.  The result of this numerical analysis is shown in \Fig{fig:cusp}, in which we plot the leading, next-to-leading and  next-to-next-to-leading order distributions, fixing $\alpha_s = 0.118$ and the number of active quarks $n_f = 5$ in QCD.  In going from leading next-to-leading order, we see that the cusp is actually softened and nothing like a discontinuous step seems to be starting to be resolved at next-to-leading order or beyond.

To understand this a bit more, we can determine the fixed-order expansion of the discontinuity of the derivative at the cusp order-by-order.  EVENT2 calculates the cross section in each color channel, so we separate out the ${\cal O}(\alpha_s^2)$ contributions in each color channel and numerically calculate the cusp.  To do this, we fit lines to the five points immediately above and below the location of the cusp, respectively, and then calculate the difference between the slopes of these lines.  With $\zcut = 0.1$, we find that this procedure determines the $\alpha_s$ expansion to be:
\begin{align}
&\frac{1}{\sigma_0}\frac{d}{d\log\rho}\left[
\frac{d\sigma^{+}}{d\log\rho}-
\frac{d\sigma^{-}}{d\log\rho}
\right]_{\substack{\rho = 2\zcut-\zcut^2 \\\zcut = 0.1}} \\
&
\hspace{2cm}\simeq -8.92252\frac{\alpha_s C_F}{2\pi} + \left(\frac{\alpha_s}{2\pi}\right)^2C_F\left(
327\,C_F -110\, C_A + 48\,n_f T_R
\right)
+\cdots\nonumber\\
&
\hspace{2cm}\simeq-1.89342\,\alpha_s + 7.6\,\alpha_s^2
+\cdots\,.\nonumber
\end{align}
In QCD, the adjoint Casimir $C_A = 3$ and $T_R = 1/2$ and we don't quote uncertainties on the ${\cal O}(\alpha_s^2)$ values as they are meant to be representative, not precise.  The next-to-leading order correction to the discontinuity to the derivative is opposite in sign to the leading-order discontinuity, resulting in a smoother distribution at higher orders.  This suggests that the description of the cusp and its resolution through higher fixed-orders converges, with no need for resummation of soft and collinear emissions around the cusp region.

Our focus here on the cusp region has been restricted to the case of mMDT grooming, or soft drop with $\beta  =0$.  This is primarily because the highest-accuracy fixed-order predictions are available for this particular groomer.  Nevertheless, some statements about the $\beta > 0$ soft drop groomers can be made.  As $\beta$ grows, the groomer weakens and ultimately provides no grooming for $\beta \to \infty$.  In this limit, there is no cusp present in the differential cross section, so we anticipate that even at leading-order, the cusp softens as $\beta$ increases.  This expectation is borne out in the analytic results of \Refs{Larkoski:2014wba,Frye:2016aiz}, for example.  With the cusp in the mMDT/$\beta = 0$ soft drop groomed mass distribution softened by inclusion of higher-orders, it is expected that the cusp in the $\beta > 0$ distributions is also softened.

\section{Factorization-Violating Contributions}\label{sec:factvio}

All-orders resummation of the groomed jet mass has been accomplished at the highest accuracy through factorization of the different components to the cross section, at leading power in the limit in which $\rho \ll \zcut \ll 1$ \cite{Frye:2016aiz}.  We won't review the factorization theorem here, and instead just point the interested reader to the original literature.  In this strongly-ordered limit in which $\rho \ll \zcut \ll1$, all emissions that remain in the jet after grooming are necessarily collinear, within an angular distance $\theta^2$ of the jet axis of
\begin{equation}
\theta^2 \lesssim \frac{\rho}{\zcut} \ll1\,,
\end{equation}
by assumptions of the factorization theorem.  Because of this effective collinear restriction, no non-global logarithms in the mass $\rho$ are present in this limit, and with mMDT grooming, all simultaneously soft and collinear divergences in the mass are also eliminated.  This significantly simplifies the structure of the emissions that can contribute to the groomed mass, hence enabling high precision resummation.

This leading-power factorization theorem can be used to predict all contributions to the cross section of the groomed heavy hemisphere mass that are enhanced by logarithms of $\rho$ and/or $\zcut$.  That is, the factorization theorem predicts the cross section to be a function of $\log \rho$ and $\log \zcut$:
\begin{equation}
\frac{d\sigma^{\rho\ll\zcut\ll 1}}{d\log\rho} \equiv \frac{d\sigma^{\rho\ll\zcut\ll 1}(\log\rho,\log\zcut)}{d\log\rho}\,,
\end{equation}
and all contributions from positive powers of $\rho$ or $\zcut$ are formally suppressed in this limit.  As we measure the cross section differential in $\rho$, we can always restrict to a region in which $\rho \ll 1$, and therefore power corrections in $\rho$ would be numerically suppressed.  However, because $\zcut$ is a fixed parameter of the groomer, the assumption of $\zcut \ll 1$ is not necessarily satisfied for any application of the groomer.  In particular, a typical value of $\zcut$ is about $0.1$, which is small, but the largest that $\zcut$ can possibly be is $0.5$, and it's not obvious that 0.1 is parametrically smaller than 0.5.  At the very least, we should assess the potential impact of finite $\zcut$ corrections to the resummation accomplished in the factorization theorem.

While we restrict our attention to grooming in perturbation theory in this paper, there are additional power corrections that arise from non-perturbative physics.  These power corrections are controlled by the ratio of the QCD scale to the jet energy scale, rather than by a finite $\zcut$ value.  However, additional calculable and perturbative finite $\zcut$ dependence can multiply non-perturbative matrix elements that lead to corrections to the groomed mass distribution \cite{Hoang:2017kmk,Hoang:2019ceu}.  For a precision prediction throughout phase space and for comparison to data, these non-perturbative corrections must be included, but we leave a complete consideration to future work.

With this goal in mind, we can express the differential cross section for the groomed heavy hemisphere mass in the regime in which $\rho\ll\zcut$, but with no restriction on the value of $\zcut$ as:
\begin{equation}
\frac{d\sigma^{\rho\ll\zcut}}{d\log\rho} = \frac{d\sigma^{\rho\ll\zcut\ll 1}}{d\log\rho} + \zcut \frac{d\sigma^{\rho\ll\zcut}_1}{d\log\rho}+ \zcut^2 \frac{d\sigma^{\rho\ll\zcut}_2}{d\log\rho}+\cdots\,,
\end{equation}
where the $\cdots$ represents terms at higher powers of $\zcut$.  The factorization theorem only describes the first term in this series in $\zcut$ and no systematic procedure has been presented as of yet to calculate the cross section coefficients of $\zcut^i$ in this series to arbitrary order in the coupling $\alpha_s$.  Further, as powers of $\zcut$ have been made explicit in this expansion, we can estimate the relative size of the power corrections in $\zcut$ to the cross section valid in the $\rho\ll\zcut\ll1$ limit.  We assume that $\rho \ll 1$ in every term on the right side, so every term should be some function of $\log \rho$.  As such, we do not expect any parametric difference between the $\rho\ll\zcut\ll 1$ term and the other cross sections, stripped of their $\zcut$ dependence.  That is, we expect
\begin{equation}
\frac{d\sigma^{\rho\ll\zcut}_i}{d\log\rho}\sim\frac{d\sigma^{\rho\ll\zcut\ll 1}}{d\log\rho}\,.
\end{equation}
Therefore, all scaling of terms in this expansion are carried by the explicit powers of $\zcut$, and so we would expect that the factorization theorem in the regime $\rho\ll\zcut \ll 1$ describes the cross section when $\rho \ll\zcut$ up to corrections of order $\zcut$:
\begin{equation}
\frac{d\sigma^{\rho\ll\zcut}}{d\log\rho} = \frac{d\sigma^{\rho\ll\zcut\ll 1}}{d\log\rho} + {\cal O}(\zcut).
\end{equation}
Concretely, if $\zcut = 0.1$, we expect the factorization theorem to correctly describe the cross section in this region up to $10\%$ corrections.

With the factorization theorem and the complete fixed-order cross section through next-to-next-to-leading order, we can test this assumption.  First, we expand the all-orders cross section of the factorization theorem in powers of $\alpha_s$ as:
\begin{equation}
\rho\frac{\rd\sigma^{\rho\ll\zcut\ll1}}{\rd\rho} = \rho\frac{d\sigma^{{(0)}\rho\ll\zcut\ll1}}{d\rho}+\rho\frac{d\sigma^{{(1)}\rho\ll\zcut\ll1}}{d\rho}+\rho\frac{d\sigma^{{(2)}\rho\ll\zcut\ll1}}{d\rho}+\cdots\,.
\end{equation}
The superscript $(n)$ denotes the term at order $\alpha_s^{n+1}$ in the limit in which $\rho \ll \zcut \ll1$.  The first three terms have been calculated and are \cite{Frye:2016aiz,Kardos:2020ppl}:
\begin{align}
\frac{2\pi}{\alpha_s}\rho\frac{d\sigma^{{(0)}\rho\ll\zcut\ll1}}{d\rho} &= -\frac{16}{3}\log\zcut-4\,,\\
\left(\frac{2\pi}{\alpha_s}\right)^2\rho\frac{d\sigma^{{(1)}\rho\ll\zcut\ll1}}{d\rho} &\simeq \left(
28.444\log^2\zcut+63.111\log\zcut+31.333
\right)\log\rho-14.222\log^3\zcut\nonumber\\
&
\hspace{1cm}-39.877\log^2\zcut-98.801\log\zcut-61.967\nonumber\\
\left(\frac{2\pi}{\alpha_s}\right)^3\rho\frac{d\sigma^{{(2)}\rho\ll\zcut\ll1}}{d\rho} &\simeq \left(
-75.85\log^3\zcut-334.22\log^2\zcut-451.70 \log\zcut-182.78
\right)\log^2\rho\nonumber\\
&
\hspace{1cm}
+\left(75.85\log^4\zcut+269.56\log^3\zcut+1008.64\log^2\zcut\right.\nonumber\\
&
\hspace{6cm}
\left.+1762.95\log\zcut+877.52
\right)\log\rho\nonumber\\
&
\hspace{1cm}-18.96\log^5\zcut-37.59\log^4\zcut-230.06\log^3\zcut\nonumber\\
&\hspace{1cm}-724.49\log^2\zcut-1641.62\log\zcut-\left(2670\pm125\right)\,.
\nonumber
\end{align}
Here we substituted explicitly the color factors of QCD ($C_F = 4/3$, $C_A = 3$, $T_R = 1/2$), and set the number of active quarks to $n_f = 5$.  As written, this is a function of $\zcut$ and so the numerical size of the terms is still obscured.  Setting $\zcut = 0.1$, the leading-power cross section is:
\begin{align}
\left.\rho\frac{d\sigma^{\rho\ll\zcut\ll1}}{d\rho}\right|_{\zcut = 0.1}&\simeq \frac{\alpha_s}{2\pi}8.28045+\left(\frac{\alpha_s}{2\pi}\right)^2\left(
36.824\log\rho + 127.807
\right)\\
&
\hspace{2cm}+\left(\frac{\alpha_s}{2\pi}\right)^3\left(
11.30\log^2\rho+1007.23\log\rho + (248\pm125)
\right)+\cdots\,.\nonumber
\end{align}

To assess the size of the finite $\zcut$ corrections order-by-order, we will calculate the fractional difference between the complete cross section in the $\rho\ll\zcut$ limit and the leading-power prediction:
\begin{equation}
\Delta^{(n)}\equiv \frac{d\sigma^{{(n)}\rho\ll\zcut}-d\sigma^{{(n)}\rho\ll\zcut\ll1}}{d\sigma^{{(n)}\rho\ll\zcut\ll1}}\,.
\end{equation}
From our earlier arguments, we expect $\Delta^{(n)} \sim \zcut$.  Starting with $n=0$, we can compare the complete leading order cross section expanded for $\rho\ll \zcut$ of \Eq{eq:lolp} to the leading-power result:
\begin{align}\label{eq:lodiff}
\Delta^{(0)}=\frac{d\sigma^{{(0)}\rho\ll\zcut}-d\sigma^{{(0)}\rho\ll\zcut\ll1}}{d\sigma^{{(0)}\rho\ll\zcut\ll1}} = 
-\frac{6\zcut+4\log(1-\zcut)}{3+4\log\zcut}\,.
\end{align}
The leading term in the numerator of this expression is indeed proportional to $\zcut$, but for $\zcut \simeq 0.1$, the denominator is substantially large.  Plugging in $\zcut = 0.1$ we find
\begin{align}
\left.\frac{d\sigma^{{(0)}\rho\ll\zcut}-d\sigma^{{(0)}\rho\ll\zcut\ll1}}{d\sigma^{{(0)}\rho\ll\zcut\ll1}}\right|_{\zcut= 0.1} \simeq 0.02875\,,
\end{align}
which is about a factor of 4 smaller than $\zcut$.  The denominator of \Eq{eq:lodiff} is logarithmic in $\zcut$ and so for small excursions varies slowly.  So, as a rule of thumb, for experimentally-relevant values of $\zcut \simeq 0.1$, we can approximate
\begin{equation}
\Delta^{(0)} \simeq \frac{\zcut}{4}\,.
\end{equation}
That is, the finite $\zcut$ contributions in the leading order cross section of the groomed heavy hemisphere mass are just few percent corrections to the leading-power prediction of the factorization theorem in the limit $\rho \ll \zcut \ll 1$.

\begin{figure}[t]
\begin{center}
\includegraphics[width=7.6cm]{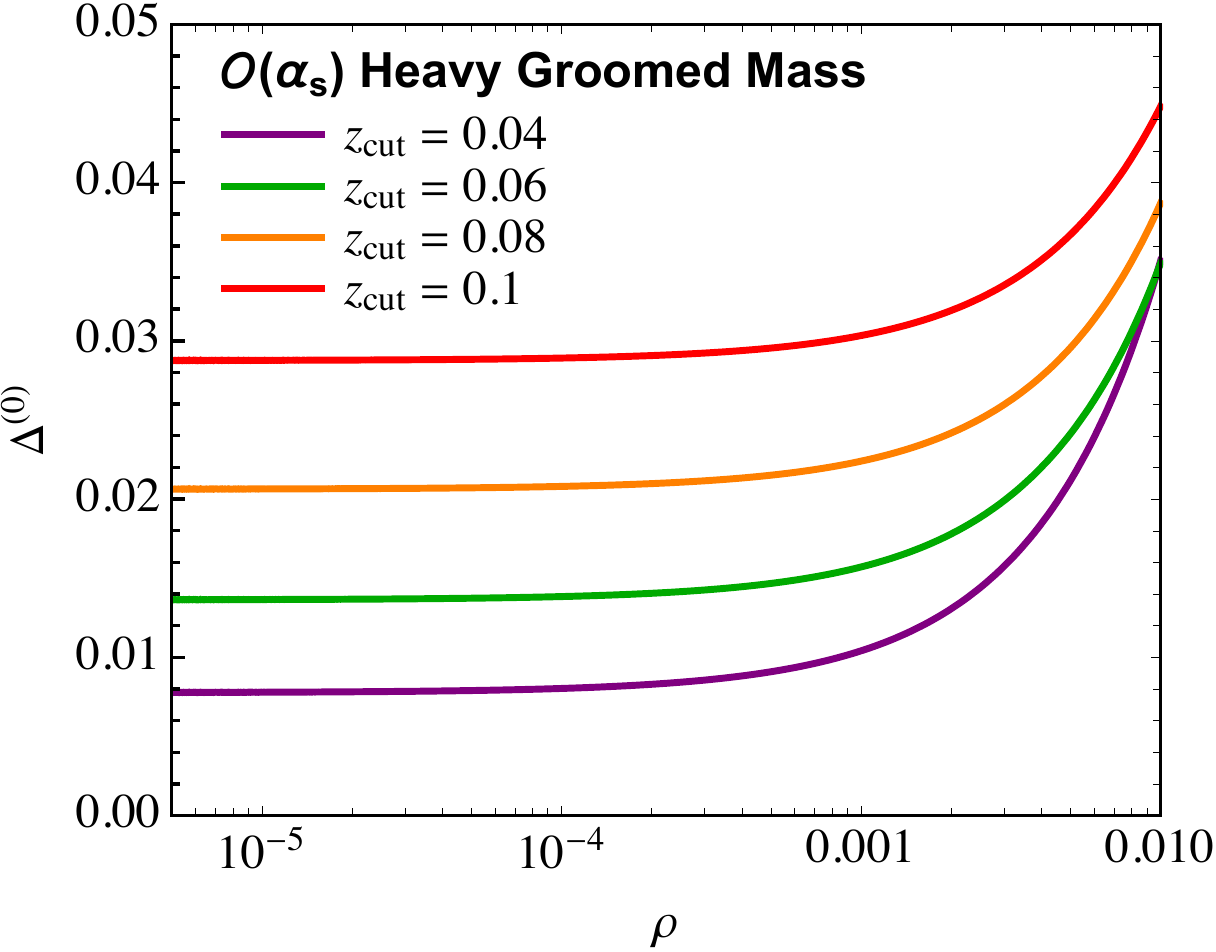} \ \ \includegraphics[width=7.6cm]{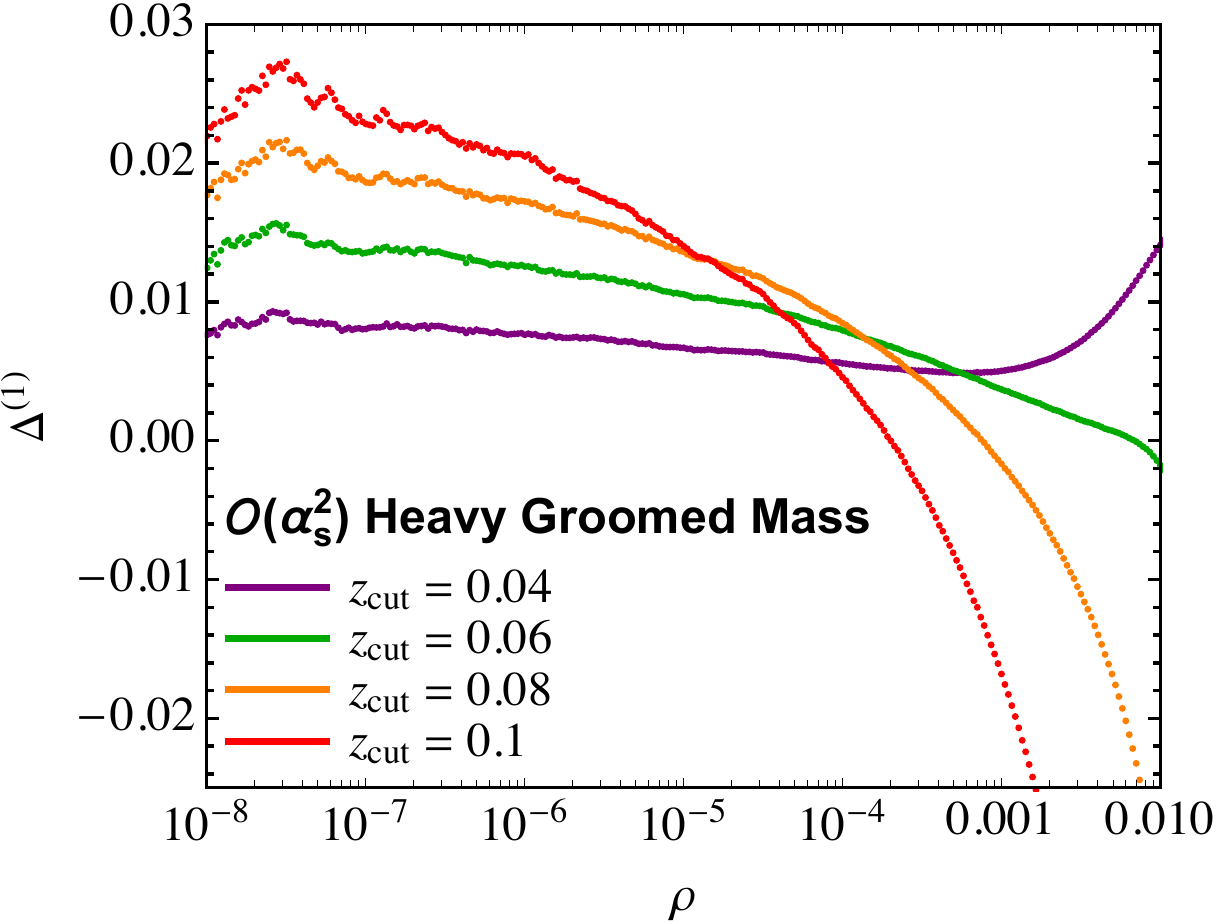}\\
\vspace{0.2cm}
\includegraphics[width=7.6cm]{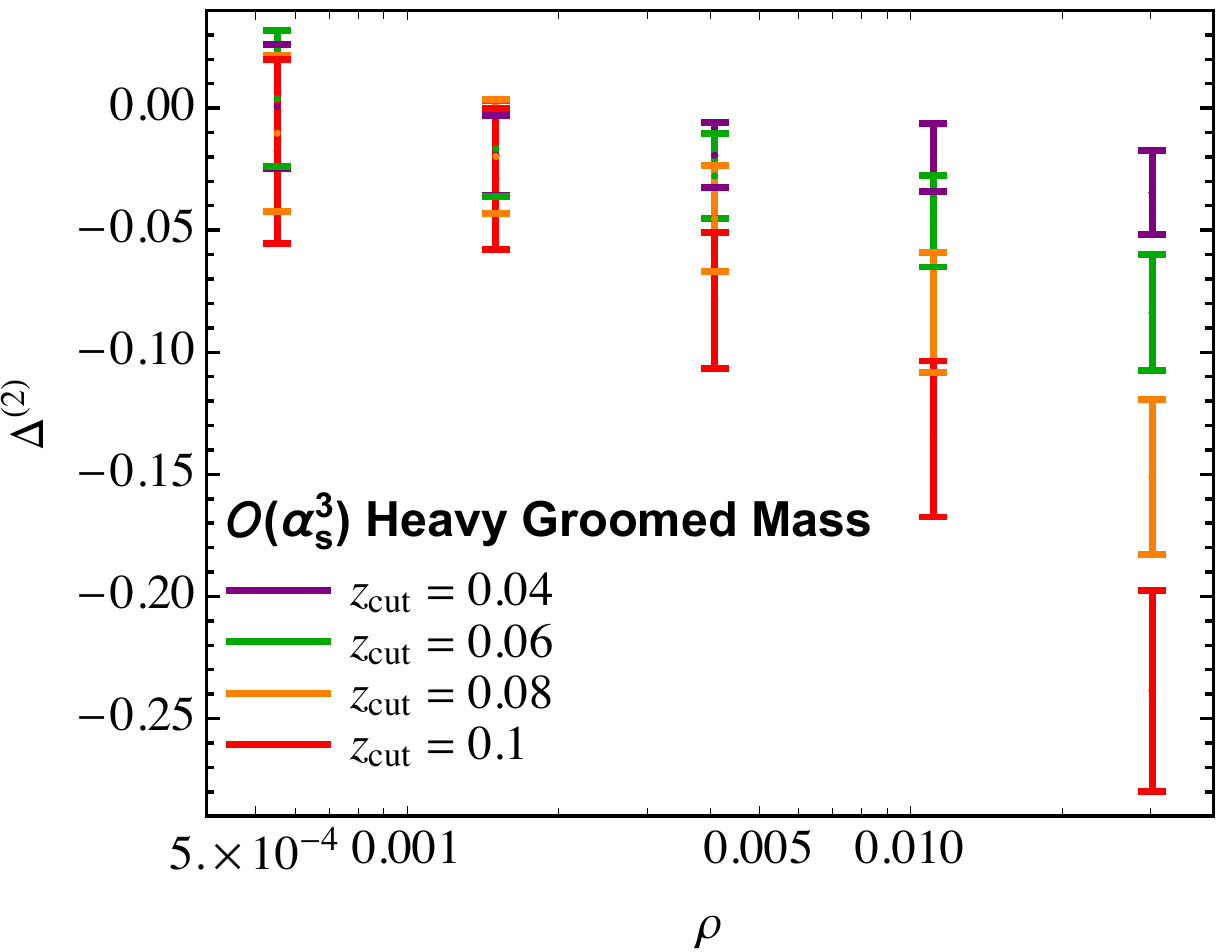}
\caption{Plots of the fractional difference $\Delta^{(n)}$ between the complete cross section and the leading-power expansion in the $\rho\ll\zcut\ll 1$ limit at ${\cal O}(\alpha_s)$ ($n=0$, upper left), ${\cal O}(\alpha_s^2)$ ($n=1$, upper right), and ${\cal O}(\alpha_s^3)$ ($n=2$, bottom).  Values of $\zcut = 0.04\,, 0.06\,, 0.08\,,0.1$ are shown.
\label{fig:facvio}}
\end{center}
\end{figure}

We extend this fractional difference comparison through ${\cal O}(\alpha_s^3)$ in \Fig{fig:facvio}.  At ${\cal O}(\alpha_s^2)$, we compare the result of the factorization theorem to the output of EVENT2, and observe that the scaling of the fractional difference is very similar to that at leading order, as $\rho\to 0$.  That is, we can also make the approximation
\begin{equation}
\Delta^{(1)} \simeq \frac{\zcut}{4}\,.
\end{equation}
At ${\cal O}(\alpha_s^3)$, we compare the result of the factorization theorem to the output of the CoLoRFulNNLO method, as tabulated in \Ref{Kardos:2020ppl}.  The bins in $\log\rho$ are large at this order and do not extend as far into the infrared as lower orders, but a similar outcome is observed.  As $\rho\to 0$, the finite $\zcut$ corrections at ${\cal O}(\alpha_s^3)$ are significantly smaller than the expected $\zcut$ size.

In a precision prediction, one must match the leading-power resummation to fixed-order for a prediction that is accurate over all of phase space.  The simplest matching procedure is additive matching in which resummed and fixed-order results are added, and their overlap is subtracted:
\begin{equation}
\frac{d\sigma^\text{(matched)}}{d\rho} = \frac{d\sigma^\text{(fixed-order)}}{d\rho}+\frac{d\sigma^\text{(resummed)}}{d\rho}-\frac{d\sigma^\text{(resummed,fo)}}{d\rho}\,.
\end{equation}
The final term represents the resummed result expanded to the order in $\alpha_s$ at which the fixed-order prediction is accurate.  If a fixed-order prediction for the groomed heavy hemisphere mass is matched to a resummed prediction in the limit that $\rho\ll\zcut\ll 1$, these results demonstrate that the fixed-order prediction will have a residual contribution to the matched cross section in the limit that $\rho \to 0$ at the order of a few percent of the resummed prediction, due to finite $\zcut$ effects.  This is at the order of, or even smaller than, estimates of theoretical uncertainties by scale variation \cite{Frye:2016aiz,Kardos:2020gty}.  With sufficiently high fixed-order matching, the uncertainty due to not resumming the finite $\zcut$ corrections that survive in the $\rho \to 0$ limit could then be accounted for within an appropriate uncertainty budget.

\section{Conclusions}\label{sec:concs}

Jet grooming, especially with mMDT or soft drop, has opened up a new precision regime in jet substructure.  The groomer introduces a new scale $\zcut$ on the jet, beyond the scale of the measurement, and that new scale both provides opportunities and challenges for precision calculations.  Because of the grooming scale in mMDT/soft drop, non-global logarithms of the jet mass $\rho$ are eliminated at small masses.  This enables an all-orders factorization theorem in the $\rho\ll\zcut\ll 1$ limit, but also produces non-analytic behavior at leading order around $\rho \sim \zcut$ and misses finite $\zcut$ corrections in the $\rho\to 0$ limit.  In this paper, we explicitly demonstrated using fixed-order codes that both of these potential issues are benign.  Unlike endpoint cusps in the thrust distribution, for example, higher-order corrections soften the cusp in the groomed mass distribution, suggesting that the region around $\rho\sim \zcut$ becomes smooth and stays continuous as higher orders are included.  In the $\rho\to 0$ limit, finite $\zcut$ corrections through ${\cal O}(\alpha_s^3)$ are actually numerically much smaller than expected, at the percent level even for typical values of $\zcut \sim 0.1$.  This level is small enough that any residual uncertainty from not resumming finite $\zcut$ corrections can be absorbed in theoretical uncertainties.

While these results demonstrate numerical control over the groomed mass distribution, it may be desirable to have a more complete analytical understanding of the features studied here.  For instance, while no non-global logarithms are present in the groomed mass distribution as $\rho\to 0$, there is a conservation of complexity.  The non-global logarithms are pushed to the $\rho\sim\zcut$ region, and may have a relationship to the physics responsible for softening the cusp.  It should be possible to construct an effective theory for small excursions away from the cusp region, and correspondingly account for soft and collinear emissions about the cusp to all orders.  Such a study would unambiguously demonstrate whether higher-order corrections do indeed smooth the cusp or not.  Though the finite $\zcut$ corrections are numerically small, they could essentially be completely eliminated by a ${\cal O}(\zcut)$ factorization theorem, for $\rho\to 0$.  For example, we expect that enumerating and factorizing all contributions that yield the first $\zcut$ corrections should be possible, as to that order there can be at most one hard emission groomed away, for example.  Accounting for these additional effects will provide an even more precise picture of groomed jets to compare to experiment.

\acknowledgments

I thank Zhongbo Kang, Kyle Lee, Xiaohui Liu, Simone Marzani and Felix Ringer for comments on the paper and Adam Kardos and Zoltan Tr\'ocs\'anyi for comments on the paper, collaboration on related work and for assistance with CoLoRFulNNLO.  This work was facilitated in part by the Portland Institute for Computational Science and its resources acquired using NSF Grant DMS 1624776.

\bibliography{sd_notes}

\end{document}